\documentclass[pra,twocolumn,floatfix,superscriptaddress]{revtex4}

\usepackage{graphicx}
\usepackage{bm}

\def\final{1} 

\ifnum\final=1  
\newcommand{\mnote}[1]{[{\small Mohammad: \bf #1}]\marginpar{*}}
\newcommand{\vnote}[1]{[{\small Vicky: \bf #1}]\marginpar{*}}
\newcommand{\sidecomment}[1]{\marginpar{\tiny #1}}
\else 
\newcommand{\mnote}[1]{}
\newcommand{\vnote}[1]{}
\newcommand{\sidecomment}[1]{}
\fi  

\newcommand{\be}{\begin{equation}}
\newcommand{\ee}{\end{equation}}
\newcommand{\ba}{\begin{eqnarray}}
\newcommand{\ea}{\end{eqnarray}}
\newcommand{\nn}{\nonumber \\}

\newcommand{\ham}{{\mathcal{H}}}

\begin{document}

      \title{First Order Quantum Phase Transition in Adiabatic Quantum Computation}

      \author{M.~H.~S.~Amin}
      \affiliation{D-Wave Systems Inc., 100-4401 Still Creek Drive,
      Burnaby, B.C., V5C 6G9, Canada}
      \author{V.~Choi}
      \affiliation{D-Wave Systems Inc., 100-4401 Still Creek Drive,
      Burnaby, B.C., V5C 6G9, Canada}
      \affiliation{Department of Computer Science,
      Virginia Tech, Falls Church, VA 22043, USA}

\begin{abstract}

We investigate the connection between local minima in the problem
Hamiltonian and first order quantum phase transitions during an
adiabatic quantum computation. We demonstrate how some properties of
the local minima can lead to an extremely small gap that is
exponentially sensitive to the Hamiltonian parameters. Using
perturbation expansion, we derive an analytical formula that can not
only predict the behavior of the gap, but also provide insight on
how to controllably vary the gap size by changing the parameters. We
show agreement with numerical calculations for a weighted maximum
independent set problem instance.
\end{abstract}

\maketitle

\section{Introduction}

Adiabatic quantum computation (AQC) was first proposed in 2000 by
Farhi {\em et al.} \cite{Farhi} as a means to solve NP-hard
optimization problems. Later, Aharonov {\em et al.}~\cite{ADKLLR04}
proved that AQC is polynomially equivalent to conventional (gate
model) quantum computation. It is also believed that AQC is more
robust against errors caused by environmental
noise~\cite{robustness,amin,lloyd-robust08}.

In AQC, the system's Hamiltonian, usually written as
 \be
 \ham(t) = [1-\lambda(t)]\ham_B + \lambda(t) \ham_P, \label{HS}
 \ee
evolves slowly with time $t$ as $\lambda(t)$ changes monotonically
from 0 to 1 within a time $t_f$. The initial Hamiltonian $\ham_B$ is
assumed to have an easily accessible ground state into which the
system is initialized, while the ground state of the final
Hamiltonian $\ham_P$ provides the solution to the problem of
interest. In order to reach the final ground state with high
fidelity, the adiabatic theorem requires $t_f \propto g_{\rm
min}^{-\delta}$, where $g_{\rm min}$ is the minimum gap between the
two lowest energy instantaneous eigenstates of $\ham$. The power
$\delta$ can be 1, 2, or possibly some other number depending on the
functional form of $\lambda(t)$ and the distribution of the higher
energy levels \cite{Farhi,Schaller,Lidar}.

Thus, in order to address the efficiency of AQC, one needs to
analyze $g_{\rm min}$, which is unfortunately as hard as solving the
original problem if computed directly. The most fundamental problem
in AQC is therefore how to bound $g_{\rm min}$ analytically. Equally
important is how to unveil the quantum evolution blackbox by
relating the the formation of $g_{\rm min}$ to the structure of the
problem, and thus obtain insights for designing efficient
algorithms.

Besides a few special cases in which spectral gaps are computed
analytically~\cite{Analytical}, all other known studies have to
resort to numerical calculations, e.g., diagonalization \cite{Exact}
or quantum Monte Carlo (QMC) techniques \cite{Young}.
Unfortunately, these methods are limited to small problem sizes (to
date, $N{<}30$ for diagonalization and up to $128$ for QMC), and
tend not to provide much insight into why an extracted minimum gap
is large or small.

It has been recognized that during the evolution an adiabatic
quantum computer may go through a quantum phase transition (QPT)
\cite{Knysh,SchutzholdQPT,Lloyd-matter08}. The transition is first
(second) order if the change in the order parameter at the
transition point is discontinuous (continuous) \cite{SchutzholdQPT}.
In this article, we investigate the effect of problem structure,
specifically the role of local minima \cite{Reichardt,LMpaper}, on
the formation of a first order QPT.

The paper is organized as follows. In the next section, we discuss
the relation between AQC, or more specifically adiabatic quantum
optimization, to quantum phase transitions. We qualitatively
describe how a first order QPT may happen during adiabatic evolution
due to the presence of a local minimum. In section III, we use
perturbation expansion to study such QPT in more detail and obtain
analytical formula for the gap position and size. In section IV, we
examine our findings via a small size example graph of the (NP-hard)
weighted maximum independent set (WMIS) problem. We show agreement
between our perturbation calculation and numerical diagonalization.
We demonstrate controllable variation of $g_{\rm min}$ by 6 orders
of magnitude by slightly (25\%) changing the Hamiltonian parameters.
At the end, we summarize our conclusions in section V.

\section{AQC and Quantum phase transitions}

Consider a transverse field Ising Hamiltonian (\ref{HS}) with
 \ba
 \ham_P {=} \sum_i h_i
 \sigma^z_i {+} \sum_{i,j} J_{ij} \sigma^z_i\sigma^z_j \, ,  \label{HP}
 \qquad
 \ham_B {=} \, \Delta \sum_i \sigma^x_i \, ,
 \ea
where $\sigma^{x,z}_i$ are Pauli matrices for the $i$-th qubit. Let
${\cal E}$ denote an energy scale that characterizes Hamiltonian
$\ham_P$ the same way as $\Delta$ characterizes $\ham_B$. (Here, we
do not specify ${\cal E}$, but only mention its existence.) In the
interpolation Hamiltonian (\ref{HS}), depending on whether $\ham_B$
or $\ham_P$ dominates, the system will be localized in one of the
two computation (in which $\sigma^z_i$ is diagonal) or Hadamard (in
which $\sigma^x_i$ is diagonal) bases, hence delocalized in the
conjugate basis. We introduce the dimensionless parameter
 \be
 \zeta = {(1{-}\lambda)\Delta \over \lambda {\cal E}}, \label{zeta}
 \ee
which provides a measure of the relative importance of $\ham_B$ and
$\ham_P$. Let $|\psi_0\rangle$ denote the instantaneous ground state
of $\ham(t)$. At $\lambda {\approx}\, 0$, $\zeta$ is very large and
$\ham_B$ is dominant making $|\psi_0\rangle$ a large superposition
in the computation basis or localized in the Hadamard basis. This is
the quantum paramagnetic phase. As $\zeta$ is lowered, around
$\zeta_c {\approx}\, 1$ (or equivalently  $\lambda_c {\approx}\,
\Delta/(\Delta {+} {\cal E})$), the dominance shifts from $\ham_B$
to $\ham_P$, which favors localization in the computation basis.

The transition from paramagnetic to ordered phase is usually via a
QPT. For a large homogeneous system with no field, QPT happens at a
well-defined critical point $\lambda_c$ and is usually continuous,
thus second order. 
The lowest-energy excitations, delocalized over the whole system,
are gapped with a gap that shrinks at $\lambda_c$, but only
polynomially in the system size N \cite{Sachdev}.
However, this scenario may be significantly altered by inhomogeneity
\cite{disordereffect}.

In the absence of any further phase transition, the minimum gap is most likely
at $\lambda_c$. Despite the localization at $\lambda_c$, $|\psi_0\rangle$ will
still be a superposition in the computation basis for $\lambda{>}\lambda_c$,
but with a much smaller number of computational states involved. As $\lambda$
is increased, the total energy of such localized state will change and at some
later point ($\lambda^*$) it may cross another localized state. The ground
state of the system will then make a sudden transition to the new state via a
discontinuous (first order) QPT \footnote{In large scale spin glasses, those
localized states may not represent two macroscopically distinct phases.
Nevertheless, we continue using the phrase ``phase transition''.}. The gap at
the transition point will be extremely small, making $\lambda^*$ the new
position of the minimum gap. The transition at $\lambda^*$ is between two
ordered phases, in contrast to the order-disorder transition at $\lambda_c$.
There may even be more than one such transition if the ground state crosses
other localized states, but all those transitions can only happen after
localization at $\lambda_c$. An important question now is what properties of
$\ham_P$ are responsible for such a first order QPT. In the next section, we
will employ perturbation expansion to answer this question.

\section{Perturbation expansion}

\begin{figure}[t]
    \centering
        \includegraphics[width=6cm]{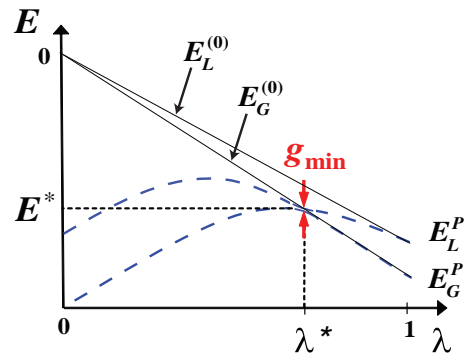}
    \caption{ Schematic diagram of crossing global and local minima.
At zeroth order perturbation, levels do not cross (solid lines).
Contribution from the second order perturbation may cause the levels
cross if the curvature of the upper level is larger than the lower
one (dashed lines). }
    \label{fig1}
\end{figure}

Let us define
 $\ham_0 {=}\, \lambda \ham_P$ and $\ham' {=}\, (1{-}\lambda)\ham_B$
as the unperturbed and perturbation Hamiltonians, respectively. We
use $\zeta$ as defined in (\ref{zeta}) as the dimensionless small
parameter. At $\zeta {=}\, 0$ ($\lambda {=} 1$), the eigenstates of
the system are eigenfunctions of $\ham_P$ (computational states).
Thus $|\psi_0\rangle$ is the global minimum of $\ham_P$ and the
lowest lying excited states are either states that are a few bit
flips away from (or neighborhood of) the global minimum, or some low
energy local minima and their corresponding neighborhoods. These two
types of states behave completely differently as $\zeta$ is
increased. Since $\ham'$ only involves $\sigma^x_i$ operators, to
every order of perturbation it flips only one qubit. Therefore, the
lowest order of perturbation that gives a nonzero off-diagonal
element $\ham_{mn} {\equiv}\, \langle m |\ham|n\rangle$ is equal to
the number of bit flips $f_{mn}$ (Hamming distance) between states
$|m\rangle$ and $|n\rangle$. This gives $\ham_{mn} {=}\,
O(\zeta^{f_{mn}})$, hence at $\lambda {\approx}\, 1$ the only
non-vanishing $\ham_{mn}$ are those between neighboring states for
which $f_{nm}$ is small. Essentially, $\ham$ becomes block diagonal
with every minimum and its neighborhood forming a cluster of states
with nonzero off-diagonal elements with each other, but vanishingly
small off-diagonal elements with states in other clusters. Upon
diagonalization of $\ham$, the new eigenstates become superpositions
of only the neighboring states. Those neighboring states never cross
due to the no-level-crossing theorem \cite{Sakurai}.
On the other hand, if as $\lambda$ is decreased, two clusters move
as a whole with respect to each other, as their energy levels cross
they create anticrossings with very small gaps of
$O(\zeta^{f_{mn}})$. The lowest of these anticrossings form a first
order QPT.

Let us now make the above observation more quantitative. At
$\lambda{=}\,1$ the Hamiltonian $\ham{=}\, \ham_P$ has a global
minimum $|G^{(0)}\rangle$ with energy $E_G^P$. Let $|L^{(0)}\rangle$
represent a low energy local minimum (not necessarily the first
excited state) of $\ham_P$ with energy $E_L^P$. For now we take both
the above states to be non-degenerate. We use perturbation expansion
to calculate the perturbed states $|\alpha {=}\, G,L\rangle$, and
their eigenvalues $E_\alpha$ at $\lambda \lesssim 1$. To the second
order perturbation, $E_\alpha {=}\, \lambda E_\alpha^P
-\chi_\alpha(1{-}\lambda)^2/\lambda$, where
 \ba
 \chi_\alpha= \sum_{n \ne \alpha} {|\langle \alpha^{(0)}| \ham_B
 | n^{(0)} \rangle|^2 \over E_n^P - E_\alpha^P}
 = {\lambda^3 \over 4}{d^2E_\alpha \over d\lambda^2}. \label{chi}
 \ea
The perturbation expansion holds as long as for all states
contributing to the sum, $E_n^P {-} E_\alpha^P \gg
(1{-}\lambda)\Delta/\lambda$. The two perturbed levels cross at (see
Fig.~\ref{fig1})
 \be
 \lambda^* = \left( 1 + \sqrt{{E_L^P - E_G^P \over
 \chi_L-\chi_G}} \right)^{-1}, \label{lambda}
 \ee
Since $E_L^P {>} E_G^P$, if $\chi_L {<} \chi_G$ then there will be
no real solution for $\lambda^*$ (up to the second order in
perturbation). Since the curvature of the energy levels, as shown in
(\ref{chi}), is proportional to $\chi_\alpha$, when $\chi_L {<}
\chi_G$ the local minimum has less curvature than the global
minimum. In that case, the lower one of the two curves in
Fig.~\ref{fig1} will be more curved than the upper one, hence will
not cross it. As a result, no first order QPT due to such a local
minimum will occur. Even if (\ref{lambda}) gives a real solution but
with $\lambda^* {<} \lambda_c$, such a solution is not acceptable
because perturbation expansion breaks down for $\lambda {<}
\lambda_c$ ($\zeta {>} 1$). Therefore, only local minima that yield
a $\lambda^* {>} \lambda_c$ may cause first order QPT. In the
absence of such local minima there wont be any first order QPT. Of
course, (\ref{lambda}) is a result of second order perturbation
expansion, therefore not accurate. In principle, one can continue to
higher orders of perturbation to find a more accurate $\lambda^*$,
but the computation becomes more complicated.

The degeneracy at the above level crossing is removed by tunneling
between the two localized states. To find $g_{\rm min}$, we need to
calculate the effective 2$\times$2 Hamiltonian $\widetilde{\ham}$ in
the subspace made of the perturbed states $|G\rangle$ and
$|L\rangle$. At the degeneracy point, $\widetilde{\ham}_{LL} {=}\,
\widetilde{\ham}_{GG} {=}\, E^*$ and the off-diagonal elements
remove the degeneracy. To the lowest order perturbation,
$g_{\rm min} {=}\,
2(\widetilde{\ham}_{LG}\widetilde{\ham}_{GL})^{1/2}$,
where
 \ba
   \widetilde{\ham}_{LG} = {(1{-}\lambda^*)^f \over \lambda^{*f-1}}
   \sum_{\{n_s\}}  {\prod_{s=1}^f
   \langle n_{s-1}^{(0)}|\ham_B|n_s^{(0)}\rangle \over
 \prod_{s=1}^{f-1} (E_G^P {-} E_{n_s}^P)}\, , \label{HLG}
 \ea
and $L {\leftrightarrow}\, G$ for $\widetilde{\ham}_{GL}$.
Here, $|n_0^{(0)}\rangle{=}\,|L^{(0)}\rangle$,
$|n_f^{(0)}\rangle{=}\,|G^{(0)}\rangle$, $f$ is the Hamming distance between
these two states, and $|n_{s\ne 0,f}^{(0)}\rangle$ cover all other states
except the above two.

If any of the local or global minima are degenerate,
$\widetilde{\ham}$ should be written in the subspace of all those
degenerate states, hence will not be a $2\times 2$ matrix. However,
if there exists a symmetry between those degenerate states, as in
the example studied below, one can still use a 2$\times$2
Hamiltonian $\widetilde{\ham}$ for the two uniform (lowest energy)
superposition states $|\alpha^{(0)}\rangle = \sum_{k\in {\cal
S}_\alpha}|k^{(0)}\rangle/\sqrt{{\cal N}_\alpha}$. Here, ${\cal
S}_\alpha$ is the subspace made of the ${\cal N}_\alpha$ degenerate
states contributing to $|\alpha^{(0)}\rangle$. In that case,
(\ref{HLG}) can be used but with the sum over all $|n_s^{(0)}\rangle
{\notin}\, S_G \cup S_L$.

As is clear from (\ref{HLG}), the gap becomes extremely small as $\lambda^*
{\to} 1$. This limit is reached, according to (\ref{lambda}), when there is a
small $E_L^P{-}E_G^P$ and/or a large $\chi_L{-}\chi_G$. The former simply means
that the local minimum should have an energy close to the global one. To
understand the latter, we notice that the denominator in the sum in (\ref{chi})
is the energy cost of a bit flip away from the local or global minima. Having a
large $\chi_L{-}\chi_G$ means that on average the cost of bit flips inside the
local minimum is smaller than inside the global minimum. In other words, there
should be more low energy excitations, close to the bottom of the local minimum
compared to the global minimum.

The above observation, although obtained from second order
perturbation, has a wider validity. In the computation basis,
$\ham_P$ is diagonal, hence plays the role of a potential energy,
while $\ham_B$ causes bit flips thus is responsible for dynamics. As
is clear from Fig.~\ref{fig1}, the unperturbed energy levels
$E_\alpha^{(0)}$ (straight lines) do not cross at any $\lambda {>}\,
0$. Thus, the curvature of the levels, which is responsible for
their crossing, is a result of the contribution of $\ham_B$.
Qualitatively speaking, for a local minimum to have a larger
curvature than (thus cross) the global minimum, it needs to gain
more contribution from $\ham_B$. This means the system should have
more dynamics, hence more freedom, within the local than the global
minimum. A simple example is when the local minimum is at the bottom
of a much wider well than the global one. Below, we construct
another example (using WMIS) in which there are several degenerate
local minima with very small energy barrier between them and a
global minimum well separated from all of those.

\section{Weighted maximum independent set example}

\begin{figure}[t]
    \centering
        \includegraphics[width=5cm]{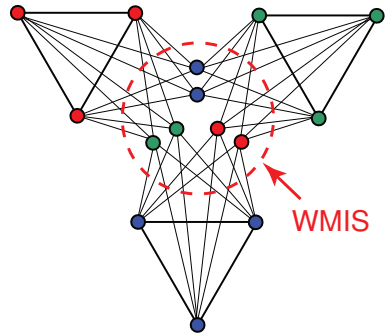}
    \caption{An example graph of WMIS problem. The
central 6 vertices have a weight $w_G$ and the 9 outer ones are
weighted $w_L$. For $w_L{<}2w_G$, the 6 central vertices make the
WMIS, while every combination of 3 vertices each from one triangle
is a smaller independent set, altogether making 27 degenerate local
minima.}
    \label{fig2}
\end{figure}


Consider a general graph ${\mathcal{G}} =
(V(\mathcal{G}),E(\mathcal{G}))$, with a set of vortices
$V(\mathcal{G})$ and a set of edges $E(\mathcal{G})$ (see
Fig.~\ref{fig2} for an example). For each vertex $i \in
V(\mathcal{G}) = \{1, \ldots, n \}$, one can associate a positive
real number $w_i$ (i.e., weight) to obtain a vertex-weighted graph.
The weighted maximum independent set problem ({\sc WMIS}) seeks to
find a ${\cal S} \subseteq V(\mathcal{G})$ such that ${\cal S}$ is
independent (i.e., each two vortices in ${\cal S}$ are not
connected) and the total {\em weight} of ${\cal S}$ ($=\sum_{i \in
{\cal S}} w_i$) is maximized. {\sc WMIS} can be solved as an Ising
problem with \cite{minor-embedding}:
 $\ham_P {=}\, \sum_{i \in V(\mathcal{G})  } h_i \sigma_i^z {+} \sum_{i,j \in
 E(\mathcal{G})} J_{ij} \sigma_i^z \sigma_j^z$,
where $h_i {=}\, \sum_{ij \in E(\mathcal{G})}J_{ij} {-} 2w_i$, with
the condition $J_{ij}>$ min$\{w_i,w_j\}$. Fig.~\ref{fig2} is an
example graph consisting $6$ (central) vertices of weight $w_G$ and
$9$ (outer) vertices of weight $w_L$, connected by edges with
$J_{ij}{=}\, J$.
For $w_L{<}\,2w_G$, the potential has a deep narrow global minimum and 27
shallow local minima separated from each other by 2 bit flips. The barrier
height between two neighboring local minima is $\delta U{=}\,4(J{-}w_L)$. Also,
one obtains
 \ba
 && E_L^P-E_G^P = 4 (6w_G - 3w_L), \nn
 && \chi_G = {1\over 4}\left({6\over w_G}+ {9\over 4J{-}w_L}\right),
 \label{Echi} \\
 && \chi_L = {1\over 4}\left({6\over 2J{-}w_G}+ {9\over w_L}+{12 \over
 J{-}w_L}\right). \nonumber
 \ea

\begin{figure}[t]
    \centering
    \includegraphics[width=8cm]{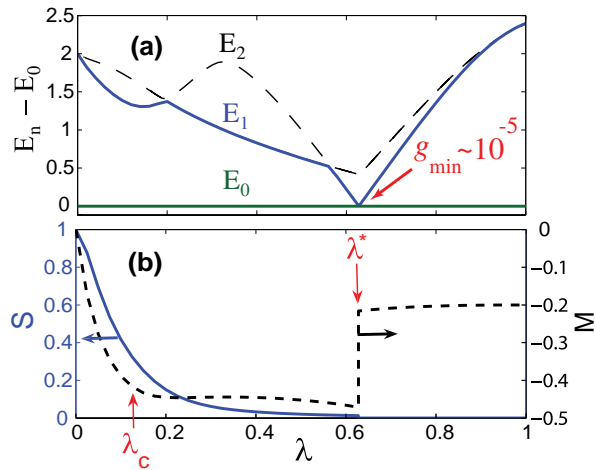}
\caption{(a) The first 3 energy levels for the example graph of
Fig.~\ref{fig2} with $J{=}\,2$, $w_G{=}\,\Delta{=}\,1$, and
$w_L{=}\,1.8$. (b) Functions $S$ and $M$, as defined in (\ref{SM}),
for the same problem.}
    \label{fig3}
\end{figure}

Figure \ref{fig3}a shows a plot of the first three energy levels for
a chosen set of parameters. The gap shows a first large minimum at
$\lambda {\approx}\, 0.1$ and a much smaller second minimum at
$\lambda {\approx}\, 0.6$. To understand what these minima
correspond to, we plot in Fig.~\ref{fig3}b two other quantities:
 \ba
 S {=} {1\over 2^{N}} [\sum_{n=1}^{2^N} |\langle \psi_0|n\rangle|\ ]^2,
 \quad M {=} {1 \over N}\sum_{i=1}^N \langle \psi_0|\sigma^z_i|\psi_0
 \rangle. \label{SM}
 \ea
$S$ provides a measure for how spread the wave function is in the
computation basis. For a wave function that is a uniform
superposition of $m$ computational states, $S{=}\,m/2^N$, i.e., the
fraction of states that participate in the superposition. $M$, on
the other hand, is the normalized total magnetization which for such
a small scale problem can play the role of an order parameter
\footnote{For large scale spin glasses $M$ is not a good order
parameter as it is nearly zero for a large number of states.}. As we
can see in Fig.~\ref{fig3}b, a significant drop of $S$ happens
around $\lambda {=}\, \lambda_c$ which coincides with the position
of the first minimum gap. Since there are only 15 qubits involved,
such a transition is not very sharp, as it would be for large scale
systems. The order parameter $M$ continuously increases in magnitude
around $\lambda_c$, but suddenly jumps to another value at
$\lambda^*$. This is a clear indication of a first order QPT at
$\lambda^*$.

\begin{figure}[t]
    \includegraphics[width=6.9cm]{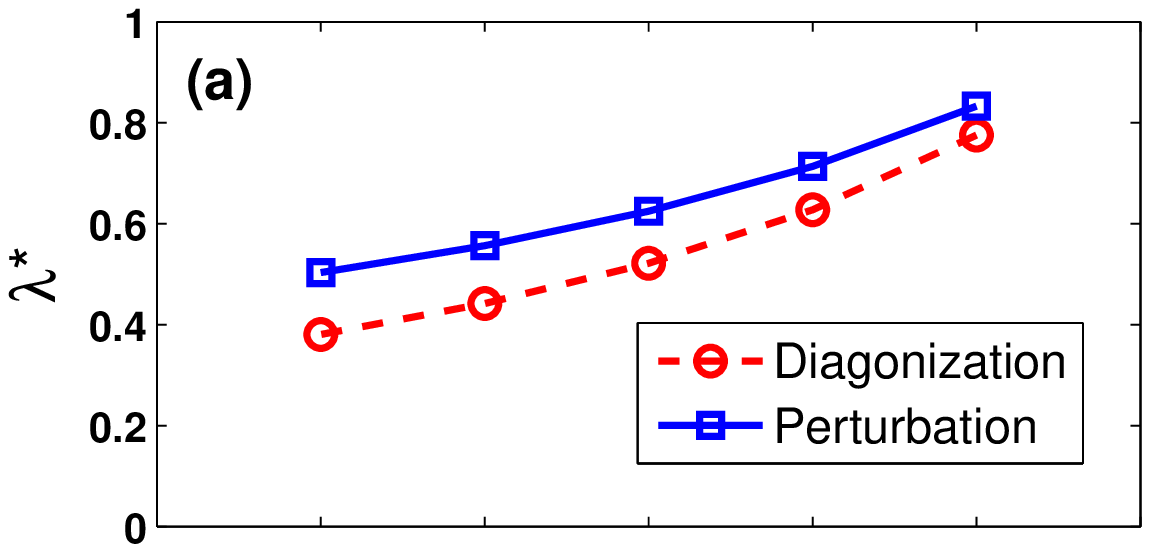}
    \includegraphics[width=6.9cm]{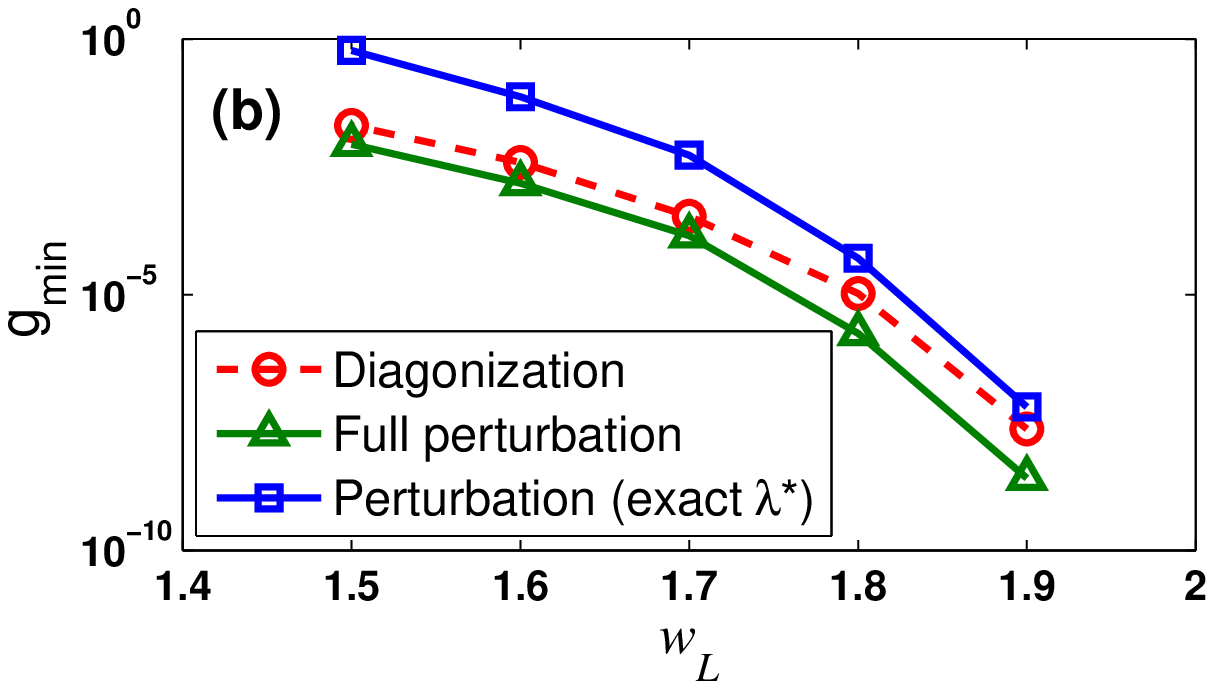}
\caption{The (a) position and (b) magnitude of the minimum gap for
the graph of Fig.~\ref{fig2} with $J{=}\,2$, $w_G{=}\,\Delta{=}\,1$.
The $x$-axis for both figures are the same.}
    \label{fig4}
\end{figure}

As $w_L {\to}\, 2$, $\delta U {\to} 0$ and the cost of bit flips
within the local minima vanishes, allowing more dynamics within the
local minima. Moreover, from (\ref{Echi}) we see $E^P_L{-}E^P_G
{\to}\, 0$. Both of these effects lead to $\lambda^* {\to} 1$ and
thereby a small $g_{\rm min}$, as discussed above.
Figure~\ref{fig4}a, plots the results of numerical diagonalization
and perturbative calculations of $\lambda^*$ versus $w_L$. Both
curves approach 1 as $w_L {\to}\, 2$.

We have also plotted $g_{\rm min}$ in Fig.~\ref{fig4}b, using both
diagonalization and perturbation. The perturbation calculations are
done at the exact $\lambda^*$ (extracted from diagonalization)
as well as its corresponding perturbative value. 
A striking point is that by changing $w_L$ by about 25\% the minimum
gap changes by 6 orders of magnitude, i.e., 12 orders of magnitude
difference in $t_f$. Such a strong dependence is indeed expected for
a first order QPT, as the tunneling amplitude between the local and
global minima is exponentially sensitive to parameters that
characterize the barrier between the minima.

\section{Conclusion}

We have shown that extremely small $g_{\rm min}$ may arise in AQC if the
anticrossing is a result of a first order QPT. Such a phase transition happens
if the final Hamiltonian possesses some low energy local minima with small cost
of bit flip within them compared to the global minimum. Our perturbative
calculation indicates that the gap depends exponentially on the Hamming
distance between the two minima involved in the phase transition. If none of
the local minima has the above properties, a first order QPT and therefore an
exponentially small minimum gap may not occur. We have also supported our
theoretical findings with an example of WMIS problem. A very small $g_{\rm
min}$, extremely sensitive to the Hamiltonian parameters, was observed as a
result of a first order QPT.

Our example, though capturing all the qualitative features of the theory, is
not generic. In many instances a minimum at $\lambda_c$ may not be visible.
Moreover, second order perturbation expansion may not be adequate to find
$\lambda^*$ with acceptable accuracy for more complicated instances.  One
should also note that a second order QPT may also lead to an exponential
dependence of $g_{\rm min}$ on the size if the Hamiltonian has random
inhomogeneities \cite{disordereffect}. Nevertheless, the physical process
leading to the first order QPT described here is general and not limited to the
presented example. Since the first arXiv appearance of this paper, similar
ideas have been employed by other groups \cite{AKR,dgosset,Young2} to a
different problem and similar results were obtained, confirming the generality
of our findings.

\section*{Acknowledgment}

We appreciate discussions with A.~Berkley, F.~Brito, S.~Han,
F.~Hamze, R.~Harris, J.~Johansson, M.~Johnson, K.~Karimi,
W.~Kaminsky, T.~Lanting, G.~Rose, and P.~Young. VC also thanks
D.~Kirkpatrick, and MHSA thanks I.~Affleck, B.~Altshuler, D.~Averin,
 S.~Haas, I.~Herbut, D. Lidar, S.~Lloyd, J.~Preskill, and W.~van Dam.

\end{document}